\documentclass[a4paper, 12pt]{article} 
\usepackage[right=1.1in,left=1.1in,top=1.1in,bottom=1.1in]{geometry}
\usepackage{graphicx} 
\usepackage{braket}
\usepackage{mathrsfs}
\usepackage[T1]{fontenc} 
\usepackage{amsmath}
\usepackage{pxfonts}
\usepackage[T1]{fontenc}
\usepackage{amssymb}
\usepackage{natbib}
\usepackage{epstopdf}
\usepackage{setspace}
\usepackage{enumitem}
\usepackage{sectsty}
\usepackage{wrapfig} 
\sectionfont{\large}
\bibpunct{(}{)}{,}{a}{}{,}
\usepackage{tikz}   
\usepackage{tikz-3dplot} 

\newcommand{\qed}{\nobreak \ifvmode \relax \else
      \ifdim\lastskip<1.5em \hskip-\lastskip
      \hskip1.5em plus0em minus0.5em \fi \nobreak
      \vrule height0.75em width0.5em depth0.25em\fi}

\usepackage{bbm}
\usepackage{MnSymbol}
\usepackage[all]{xy}

\usepackage{xcolor,colortbl,array,amssymb}

   \usepackage{braket}
   \usepackage{amssymb}
\usepackage{tcolorbox}

\makeatletter

\title{ \Large{From Time Asymmetry to Quantum Entanglement: }\\ \Large{The Humean Unification }  
} 

\author{Eddy Keming Chen\thanks{Department of Philosophy,  University of California, San Diego, 9500 Gilman Dr, La Jolla, CA 92093-0119. Website: www.eddykemingchen.net. Email: eddykemingchen@ucsd.edu} }

\date{Forthcoming in \emph{No\^us} \\  \vspace{10pt} Penultimate Version of \today} 

\begin{document}
\bibliographystyle{apalike}

\maketitle 



\begin{abstract}

Two of the most difficult problems in the foundations of
physics are (1) what gives rise to the arrow of time and (2) what the ontology
of quantum mechanics is. I propose a unified `Humean' solution to the two problems. Humeanism allows us to incorporate the Past Hypothesis and the Statistical Postulate into the best system, which we then use to simplify the quantum state of the universe. This enables us to confer the nomological status to the quantum state in a way that adds no significant complexity to the best system and solves the ``supervenient-kind problem'' facing the original version of the Past Hypothesis. We call the resultant theory  the \textit{Humean unification}. It provides a unified explanation of time asymmetry and quantum entanglement. On this theory, what gives rise to time's arrow is also responsible for quantum phenomena.  The new theory  has a separable mosaic, a best system that is simple and non-vague, less tension between quantum mechanics and special relativity, and  a higher degree of theoretical and dynamical unity.  The Humean unification leads to new insights that can be useful to Humeans and non-Humeans alike.

\end{abstract}

\hspace*{3,6mm}\textit{Keywords:  arrows of time, quantum entanglement, wave function, density matrix, cosmology, Humean supervenience, laws of nature, fundamentality, Best System Account, the Past Hypothesis, Statistical Postulate, the Mentaculus, the Wentaculus, objective probabilities, separability, narratability, nomic vagueness, Lorentz invariance}     

\newpage

\begingroup
\singlespacing
\tableofcontents
\endgroup




\nocite{feynman2011feynman, AlbertLPT, loewer2004david, lebowitz2008time, goldstein2001boltzmann, durr2005role, ney2013wave, loewer2016mentaculus, LewisPP2, sep-time-thermo, north2011time}

\section{Introduction}

Two of the most puzzling phenomena in nature are time asymmetry and quantum entanglement. They have played important roles in the development of contemporary physics. The investigation of time asymmetry started a rigorous  discipline of statistical mechanics with applications to many domains. The research into  quantum entanglement  produced  profound  insights about the nature of quantum mechanics and  technological advances in quantum information and cryptography. 

In philosophy of science, both problems are treated as useful data for evaluating leading theories about laws, chances, and ontology. They frequently come up in debates about Humeanism vs. anti-Humeanism in the metaphysics of science, serving as important case studies regarding questions such as whether the fundamental ontology is separable, whether laws supervene on the material ontology, and whether we should allow fundamental laws about initial conditions and ``deterministic chances.''

 So far, however,  the problems of time asymmetry and quantum entanglement have largely been treated as independent issues in the foundations of physics and philosophy of science.  Humeans have offered ingenious solutions to the problems by conferring nomological status to the Past Hypothesis (PH), a promising explanation for the arrow of time in our patch of the universe, and (recently) to the quantum wave function, which is responsible for the phenomena of quantum entanglement.  However, conferring  nomological status is not always easy and could lead to  tensions with other things Humeans may believe about laws of nature. For example, can PH be a fundamental (Humean) law even if it is stated in a non-fundamental language, as an infinitely long disjunction, or with vague terms? Can the wave function be considered nomological if it is extremely complex and perhaps more complex than the mosaic it aims to summarize? Independent answers have been proposed but they seem to require further modifications of the Humean framework, which may not be fully satisfactory. 
 
 The purpose of this paper is to focus on the connections between the two problems and show that they are deeply related such as to permit a unified treatment in the Humean framework. The unification in the Humean framework shows that what is responsible for time's arrow can  also be responsible for the non-separable phenomena in nature. I do this in three steps. First, I propose a new theory of quantum statistical mechanics. Second,  I use PH to select a natural initial quantum state of the universe. Third,  I use the nomological status of PH to argue for the nomological status of the quantum state of the universe. I call the general strategy the \emph{Humean unification}. I show that it leads to not only novel solutions to both problems but also new insights about the relationship between Humeanism and foundations of physics.  The success of Humean unification suggests that,  when solving difficult problems in philosophy of science,  it can be tremendously useful to take on the perspective of Humeanism, even if one is not a Humean.  The Humean unification strategy can be adapted for certain non-Humean ``nomic'' interpretation of the quantum state and the Past Hypothesis. 

 I proceed as follows. In \S2, I review the problems of time asymmetry and quantum entanglement in more detail and discuss their relevance to the Humean framework. In \S3, I review the Mentaculus theory, a promising and concrete theory of quantum statistical mechanics, and I construct a new theory called the \emph{Wentaculus} that makes central use of density matrices and a new law called the  \emph{Initial Projection Hypothesis} that replaces PH. In \S4, I ``Humeanize'' the Wentaculus by arguing that the initial quantum state of the universe described by the Initial Projection Hypothesis can be interpreted nomologically rather than ontologically, which leads to a unified treatment of time asymmetry and quantum entanglement.  In \S5, I discuss the fruits of Humean unification. In \S6, I compare and contrast Humean unification to other related proposals that focus on only one of the two problems.\footnote{In this paper, I make use of ideas and methods from the metaphysics of science, philosophy of physics, and mathematical physics. I do not worry too much about whether the ideas are purely philosophical or scientific. A precise disciplinary boundary here is difficult to draw. Indeed,  I  welcome the possibility that  philosophical ideas can lead to new theoretical possibilities in foundations of physics and vice versa. }

The issues  discussed here have ramifications that go beyond the plausibility of Humeanism. By choosing a \emph{natural} initial quantum state of the universe, our theory provides novel insights about  the foundations of quantum statistical mechanics.\footnote{This paper is the second part of a project called ``Time's Arrow in a Quantum Universe.'' For other related papers in the project, see \cite{chen2018IPH, chen2019quantum1, chen2018NV, chen2018valia}. }

\section{Problems of Time Asymmetry and Quantum Entanglement}

\subsection{The Original Problems}
In this section, I discuss the original problems of time asymmetry and quantum entanglement as well as further problems they give rise to.

The first problem can be stated as follows:
\begin{description}
\item[A. The Problem of Time Asymmetry:] Why is the world temporally asymmetric when the fundamental dynamical laws are (essentially) symmetric in time? 
\end{description}
Time asymmetry is widespread in nature:   ice cubes melt in a cup of hot tea but do not spontaneously form in it; gas expands in a box but does not spontaneously contract; wine glasses break into pieces but the broken pieces do not spontaneously form wine glasses. In the language of thermodynamics,  (isolated) physical systems (typically) evolve from states of lower entropy  to states of higher entropy; but not the other way around. The phenomena are encapsulated in the Second Law of Thermodynamics: (isolated) physical systems (typically) do not decrease in entropy. However, the fundamental dynamical laws of physics, such as the Newtonian equation of motion, the Schr\"odinger equation,  the Dirac equation, and Einstein field equations are  (essentially) symmetric in time. They allow ice cubes to decrease in size and to increase in size, gas molecules to expand and to contract, and so on. They allow (isolated) physical systems to increase in entropy as well as to decrease in entropy. 

It has been argued that the origin of time asymmetry in our universe lies in a low-entropy boundary condition, now called the \emph{Past Hypothesis} (PH).\footnote{ PH was originally suggested in \citep{boltzmann2012lectures}[1898] (although he seems to favor another postulate that can be called the \textit{Fluctuation Hypothesis}) and  discussed in \cite{feynman2017character}[1965]. For recent discussions, see \cite{albert2000time}, \cite{loewer2016mentaculus}, \cite{callender2004measures, sep-time-thermo},  \cite{north2011time}, \cite{lebowitz2008time},  \cite{goldstein2001boltzmann}, and \cite{goldstein2019gibbs}. The  phrase `Past Hypothesis' is coined by  Albert (2000). See \cite{earman2006past} for worries about  PH. See \cite{goldstein2016hypothesis} for a discussion about the possibility, and some recent examples, of explaining the arrow of time without PH. } According to PH,  the universe ``started'' in a state of extremely low entropy. Starting from that state, \emph{most likely} the universe will evolve according to the fundamental physical laws into higher entropy states, giving rise to the temporal asymmetry we observe. We add the probabilistic qualifier ``most likely'' because there exist some initial conditions compatible with PH that will go to lower-entropy states in the future. To provide a reason for neglecting those anti-entropic states, we postulate a uniform probability distribution according to which they are extremely unlikely. That distribution is specified by the \emph{Statistical Postulate} (SP). \cite{loewer2012two} dubs the package of postulates---the dynamical laws, PH, and SP---the Mentaculus.  

However, PH and SP give rise to difficult conceptual issues.
 Since they play a crucial role in explaining time asymmetry and the Second Law, and since they are incredibly simple, it has been argued that PH is a fundamental law of nature\footnote{ Suggestions that PH is an additional law of nature can be found in \cite{feynman2017character, albert2000time, goldstein2001boltzmann, callender2004measures, loewer2012two, goldstein2019gibbs} and \cite{chen2020harvard}. The inference that PH holds the elite status of a \emph{fundamental} law is  based on the fact that it does not seem to be derived from anything else.  In this paper, due to lack of space, I set aside the interesting possibility raised by \cite{carroll2004spontaneous}. } and the Statistical Postulate underwrites objective probabilities. But how can PH be a law of nature if it is a (macroscopic) boundary condition? And how can the initial probability distribution be objective if the laws are deterministic? 
\begin{enumerate}
\item[A1.] \textbf{The Status of PH:} How can PH be a fundamental law of nature if it is a (macroscopic) boundary condition?
\item[A2.] \textbf{The Status of the Statistical Postulate:} How can the initial probability distribution be objective if the laws are deterministic? 
\end{enumerate}

The second and seemingly unrelated problem is as follows:
\begin{description}
\item[B. The Problem of Quantum Entanglement:] What is the nature of quantum entanglement?
\end{description}
Quantum mechanics is one of the most  successful physical theories.  But it presents numerous conceptual puzzles. At the heart of them is the phenomenon of quantum entanglement. Quantum entanglement is a property of the quantum state, which is standardly represented by a wave function $\psi$. Two systems $A$ and $B$ are entangled when their joint state $\psi_{AB}$ cannot be decomposed as a product of their individual states. We have good reasons to be \emph{realist} about the quantum state.\footnote{See \cite{chen2019realism} for a survey of realism about the quantum state.} So we may have to postulate the quantum state as part of the fundamental ontology. That would make the fundamental ontology  \emph{non-separable}: the fundamental state of the world is not determined by the states of its parts. Quantum entanglement is a kind of  \emph{holism}.\footnote{See \cite{miller2016quantum} for more discussions about quantum holism.} However, this is not the only surprising consequence of quantum entanglement. David \cite{albert2015after} has shown that if quantum entanglement is among the fundamental facts, i.e. in the mosaic, then Lorentz invariance of special relativity would conflict with a very natural principle called \emph{narratability}: the full history of the world can be narrated in a single temporal sequence, and other ways of narrating it will be its geometrical transformations (e.g. by Lorentz transformations). The conflict could be a problem for Everettian  (and some GRW-type) theories that aspire to be (fully) Lorentz invariant. 
\begin{enumerate}
\item[B1.] \textbf{The Problem of Non-Separability:} The state of the world is not determined by the states of its parts. 
\item[B2.] \textbf{The Conflict between Lorentz Invariance and Narratability:} If quantum entanglement is in the mosaic, then Lorentz invariance conflicts with narratability. 
\end{enumerate}

\subsection{Relevance to Humeanism}

The problem of time asymmetry and the problem of quantum entanglement have been much discussed in foundations of physics and metaphysics of science. Both problems have come up in the discussions of Humeanism: the first problem has been used to support Humeanism and the second one against Humeanism. They also provide inspirations for various  modifications of Humeanism. These include: \cite{LoewerHS, cohen2009better, callender2010special, miller2013quantum, esfeld2014quantum, bhogal2015humean, callender2015one, albert2015after, miller2016quantum},  \cite{esfeld2017minimalist}, among others. 

So far, the two problems have been treated independently. Interestingly, both  solutions have something to do with laws of nature. However, they face certain \emph{prima facie} problems, which I discuss below. 

The Humean\footnote{Whether the contemporary Humean position in the metaphysics of science represents the historical Hume is a controversial question.  See, for example, \cite{strawson2015humeanism}. } framework in the metaphysics of science can be summarized as follows:
\begin{itemize}
	\item Humean Mosaic: the fundamental physical ontology is a separable mosaic. In the words of \cite{LewisPP2}, it consists in ``local matters of particular fact, just one little thing after another.'' 
	\item Best System Account of Lawhood: the fundamental laws are the axioms of the summary that best balances a host of theoretical virtues such as simplicity and strength. 
\end{itemize}
Humean supervenience is the thesis that the fundamental ontology is the Humean mosaic  and all else (including laws) supervene on that. \cite{loewer2001determinism, loewer2004david} suggests that we should allow the best system to admit objective probabilities even when the laws are deterministic, so long as admitting them makes the system more informative without adding too much complexity.  This is an important modification of the original Humean framework, but it is arguably continuous with the Mill-Ramsey-Lewis account. This modification helps us better understand the success of statistical mechanics, where objective deterministic probabilities  play a central role.\footnote{For a different perspective, see \cite{schaffer2007deterministic}.} For the rest of this paper, I will adopt the modified Humean framework as the  starting point. 

On the one hand, in response to the problem of time asymmetry, it has been argued\footnote{See, for example, \cite{callender2004measures} and \cite{loewer2012two}. } that the (modified) Humean framework can solve the problem and the worries raised in A1 and A2. It is highly plausible that PH and SP belong to the Humean best summary, as they vastly increase the  informativeness of the system  without adding much complexity.  On the (modified) Humean account, PH and  SP are as nomic as the dynamical equations of motion. This is true despite the fact that PH describes a boundary condition and the fundamental dynamical laws may be deterministic. 

However, new problems arise as we consider the character of the two new Humean laws.  PH faces a ``language problem.''\footnote{Some may respond that we can just replace the PH and SP by specifying a single probability distribution on the state space that does not suffer from the language problem described here. But it is plausible that, in the standard framework, the simplest way to specify SP is to first specify the initial macrostate (using something like PH) which will then serve as the ``support'' of the probability distribution. New possibilities will be available in the Wentaculus framework. See \S3.2. } Adapting the terminology of \cite{cohen2009better}, we can also call it the ``supervenient-kind problem'': terms such as entropy is obviously not fundamental, and as such PH may not be fit to be a fundamental law \emph{if the axioms of the best system need to be stated entirely in fundamental physical terms}.\footnote{This is essentially Lewis's insistence that the terms of the best system must refer to ``perfectly natural properties,'' or the fundamental properties picked out by fundamental physics. See \cite{SiderWBW} for a similar proposal for ``structural'' and ``joint-carving'' properties. What these amount to and whether this requirement is tenable is a controversial issue. But it is important to remember that the notion of naturalness plays an important role in response to the trivialization problem of $(x)Fx$ and the new riddle of induction.  } They note that if we flesh out ``low entropy'' in terms of the microlanguage, it will be an infinitely long disjunction of microstates which does not seem to be simple at all. (The supervenient-kind problem is part of the motivation for ``language-relativization'' in the Better Best System Account, which is arguably a radical departure from the original Humean framework.)

In fact, the language problem runs deeper: not only is an infinite disjunction probably too long to be an axiom of the best system, but also  the macroscopic terms such as entropy do not  correspond  to exactly one set of disjuncts.\footnote{Since PH and SP are considered to be fundamental laws of nature, they are examples of what I call \emph{nomic vagueness}. For more detail, see   \cite{chen2018NV} and \cite{ChenNS2020}.} Given the inherent vagueness in connecting  the macroscopic and the microscopic,  it is plausible that there will be borderline cases of whether some microstates fall under the allowed range of states described by PH. Macrostates have vague boundaries. Any precise boundary would seem artificial and arbitrary.  In so far as it is desirable to keep all Humean laws exact, the vagueness of the Past Hypothesis is another reason to reconsider the Mentaculus.

 On the modified Humean framework, SP is as objective as  the postulates of quantum-mechanical probabilities. If one qualifies as chance, the other does too.  The two sources of chance could be related. It would be desirable if we can unify the two or reduce one to the other. 

On the other hand, the problem of quantum entanglement is \emph{prima facie} threatening to Humeanism.  First,  \cite{MaudlinMWP} suggest that B1 is a problem for Humeanism, as the entanglement relations would make the mosaic non-separable.\footnote{See \cite{teller1986relational} for a related argument for ``relational holism.''}   Second, if we would like to keep Lorentz invariance (e.g. for Everettian theories) in a non-separable mosaic, we would have to sacrifice narratability. This is undesirable. Since Lorentz transformations does not fully preserve the quantum-mechanical data as argued by Albert (2015), in order to tell the story of the mosaic in a temporal sequence, we would have to specify the quantum state not just along one foliation but along all foliations of space-like hypersurfaces. Describing the Humean mosaic temporally will become infinitely more complex---a potentially undesirable result. 

A promising response to B1 is also a ``nomic'' strategy:  the quantum state (represented by a wave function) is removed from the mosaic and placed into the best system. In quantum theories that postulate physical things in addition to  the quantum state (such as Bohmian mechanics, GRW spontaneous collapse theories with a matter density ontology or a flash ontology, and Everettian quantum theory with a matter density ontology; see \cite{allori2013primitive}), it may be tempting to think that the quantum state is part of the summary of the local ontology consisting in particles, matter density, or flashes in physical space. However, the quantum state may be too complex to be nomological. In fact, the typical quantum wave function is highly detailed and complicated as a function on configuration space. In response to the complexity worry, one may follow \cite{goldstein1996bohmian, goldstein2001quantum, goldstein2013reality}
 to connect the nomological interpretation to the Wheeler-DeWitt equation in quantum gravity. As a solution to that equation, the universal wave function must be time-independent and thus \emph{may} be simple.  However, this is by no means guaranteed. In any case,  defenders of Humeanism would be  wise  to also explore other possibilities, especially since it is not clear whether the Wheeler-DeWitt equation will survive future development in quantum gravity. 

Nevertheless, the nomological interpretation seems promising and on the right track, especially since a successful nomological interpretation can solve both problems---B1 and B2---at the same time. If the entanglement relations are not in the mosaic, then it can satisfy separability, narratability, as well as Lorentz invariance (for theories with such an aspiration). This is in contrast to the high-dimensional Humean interpretation of the wave function developed by \cite{LoewerHS}. 

So far, the treatments of the two problems are largely independent. But the two problems are in fact deeply related. The Humean unification will take advantage of their connections. I use PH to simplify the quantum state (by choosing a natural, simple, unique, objective, but mixed quantum state), so that we can solve  B1 and B2 without making the law system overly complicated . I then use the chosen quantum state to connect the initial low-entropy macrostate to the microdynamics, providing a solution to the supervenient-kind problem and the vagueness problem.

\section{Towards A New Theory}

In this section, I first review the standard account of quantum mechanics in a time asymmetric universe. For concreteness, I focus on the quantum Mentaculus, a neo-Boltzmannian account of quantum statistical mechanics. Next, I propose an alternative account called  the \emph{ Wentaculus.}  It replaces the universal wave function with a (mixed-state) universal density matrix, the pure-state dynamics with mixed-state dynamics, and PH with the Initial Projection Hypothesis. (Here I focus on the conceptual ideas as much as possible, leaving most mathematical details to the footnotes.) 

\subsection{The Mentaculus}

Understanding the physical world requires simple explanations for all the nomic regularities. As  discussed in \S2, many regularities that are most familiar to us, such as ice melting and smoke dispersing, are time-asymmetric. A large class of these time-asymmetric phenomena can be understood as entropic asymmetries in time: the past events have lower entropies than future events. To give a full account of such asymmetries, we can  postulate low-entropy boundary conditions and probability distributions in addition to the fundamental dynamical equations. The field of statistical mechanics has devoted considerable energy in justifying the conjecture that these postulates can explain a typical monotonic increase in entropy. Here I focus on the proposal of \cite{albert2000time} and \cite{loewer2012two, loewer2016mentaculus} that is inspired by  Boltzmann's approach to statistical mechanics \citep{goldstein2001boltzmann}:

\begin{tcolorbox}
\centerline{\textbf{The Classical Mentaculus}}
\begin{enumerate}
\item \textbf{Fundamental Dynamical Laws (FDL):} the classical microstate of the universe is represented by a point in phase space\footnote{The phase space is the $6N$-dimensional state space for a classical system with $N$ point particles that move in physical space.} (encoding the positions and momenta of all particles in the universe) that obeys $F=ma$, where $F$ encodes the classical interactions. 
 \item \textbf{The Past Hypothesis (PH)}: at a temporal boundary of the universe, the microstate of the universe lies inside $M_0$, a low-entropy macrostate that, given a choice of C-parameters,\footnote{The C-parameters are certain conventional choices---the coarse-graining variables---that connect the macrostates to sets of microstates.} corresponds to a small-volume set of points on phase space that are macroscopically similar. 
\item \textbf{The Statistical Postulate (SP)}: given the macrostate $M_0$, we postulate a uniform probability distribution\footnote{The uniform probability distribution here is with respect to the canonical Lebesgue measure on phase space. } over the microstates compatible with $M_0$.
\end{enumerate}
\end{tcolorbox}

This is  the classical-mechanical version of the Mentaculus theory. It is a version of the Boltzmannian account of classical statistical mechanics. The Mentaculus goes beyond the usual Boltzmannian approach, which is based on a notion of typicality instead of direct probability, as the Mentaculus specifies a particular low-entropy macrostate $M_0$ and a particular probability distribution (the uniform one).  The exact differences do not matter for our purposes here. Most theorems and conjectures in statistical mechanics apply to it just as well as they apply to weaker versions of the PH and SP. I chose to focus on the Mentaculus  as a representative of a standard way of thinking about time's arrow in a classical universe. 

Next, we  move to quantum statistical mechanics. Let us consider how to extend the classical Mentaculus. The key will be to replace the classical state space (phase space) with the quantum state space---the Hilbert space---and to reformulate Boltzmannian statistical mechanics in terms of resources in Hilbert space. Here we can follow the suggestions of \cite{albert2000time}\S7 and the mathematical framework of \cite{goldstein2010approach}. 

\begin{tcolorbox}
\centerline{\textbf{The Quantum Mentaculus}}
\begin{enumerate}
\item \textbf{Fundamental Dynamical Laws (FDL):} the quantum microstate of the universe is represented by a wave function $\Psi$ that obeys the Schr\"odinger equation $ i\hbar \frac{\partial \psi}{\partial t} = \hat{H} \psi$, where $\hat{H}$ encodes the fundamental interactions in quantum mechanics. 
\item \textbf{The Past Hypothesis (PH)}: at a temporal boundary of the universe, the wave function $\Psi_0$ of the universe lies inside a low-entropy macrostate that, given a choice of C-parameters,\footnote{In addition to the ones mentioned in the classical Mentaculus, the C-parameters here also include conventional choices about the cut-off threshold of quantum state macrostate inclusion. }  corresponds to $\mathscr{H}_{PH}$, a low-dimensional subspace of the total Hilbert space. 
\item \textbf{The Statistical Postulate (SP)}: given the subspace $\mathscr{H}_{PH}$, we postulate a uniform probability distribution\footnote{The uniform probability distribution is with respect to the surface area measure on the unit sphere of $\mathscr{H}_{PH}$. } over the wave functions compatible with $\mathscr{H}_{PH}$.
\end{enumerate}
\end{tcolorbox}

The quantum Mentaculus is a concrete version of a standard way of thinking about time's arrow in a quantum universe. Given this setup, the aim is to show that typical universal wave functions compatible with these postulates will evolve in such a way that most subsystems increase in entropy. There have been recent  results that are \emph{highly suggestive} along this direction.\footnote{For example, see \cite{goldstein2006canonical, goldstein2010approach, goldstein2010normal}, \cite{tasaki2016typicality}, and  references therein.   } 

Let us provide some explanations of the quantum Mentaculus. First, the quantum microstate of the universe is represented by a wave function $\Psi$. It corresponds to a unit-length vector in Hilbert space. The Hilbert space is an infinite dimensional state space for quantum theory. But a slightly more perspicuous picture of the wave function is to think of it as a function on the configuration space $\mathbb{R}^{3N}$. The configuration space is analogous to the phase space in classical mechanics except that it has  $3N$ dimensions instead of $6N$ dimensions (where $N$ is the number of particles in the universe), and each point in the configuration space represents a possible configuration of particles in physical space in terms of their locations only. The wave function assigns values to every point in configuration space. How to interpret the wave function is a central question in the foundations of quantum mechanics. 

But even before  engaging in the philosophical questions about the interpretation of the wave function, it is important to realize there is a scientific question at the heart of quantum theory: what is the dynamics of quantum mechanics? The wave function changes over time and obeys the Schr\"odinger equation. Since the wave functions can superpose into other wave functions, and  since the Schr\"odinger equation is linear, we encounter the notorious \emph{quantum measurement problem}, about which we will return to shortly. 

Second, to account for the temporal asymmetry of entropy, we introduce a low-entropy boundary condition---PH. It rules out the overwhelming majority of initial wave functions in the Hilbert space, leaving a small class of wave functions that have low entropy. They all lie inside a special subspace $\mathscr{H}_{PH}$ of the total Hilbert space.  PH says that the actual initial wave function of the universe is inside  $\mathscr{H}_{PH}$. The PH subspace $\mathscr{H}_{PH}$ has very low entropy. In classical statistical mechanics, Boltzmann  entropy of a phase point is proportional to the logarithm of the volume of the macrostate that includes the phase point. Analogously,  Boltzmann entropy of a wave function is proportional to the logarithm of the dimension of the subspace it (almost entirely) belongs.\footnote{See \cite{goldstein2010approach} for more rigorous definitions.} Hence, $\mathscr{H}_{PH}$ is a low-dimensional subspace. 

Third, to make it overwhelmingly likely that the initial wave function is entropic, i.e. evolves to higher-entropy states,  we introduce the quantum version of the Statistical Postulate (SP). It provides a uniform probability distribution over the initial wave functions in the subspace. Because of time-reversal invariance, it is plausible that there exist an infinity of ``bad'' wave functions that are anti-entropic (i.e. evolve to lower entropy). But the uniform probability distribution assigns  much lower weight on them than on the entropic wave functions.\footnote{Since the wave functions have to be normalized, they form a unit sphere in the subspace.  The SP probability distribution is supported on the unit sphere $\mathscr{S}(\mathscr{H}_{PH})$.} 

These three postulates make up the quantum version of the Mentaculus. However, the Mentaculus cannot be the entire story of quantum mechanics in a time asymmetric universe. As  mentioned before,  quantum mechanics itself faces the measurement problem. It seems that the Schr\"odinger evolution of the wave function is interrupted by sudden collapses. The wave function typically evolves into superpositions of macrostates, such as the cat being alive and the cat being dead. This can be represented by wave functions on the configuration space with disjoint macroscopic supports $X$ and $Y$. During measurements, which are not precisely defined processes in the standard theory, the wave function  undergoes random collapses. The probability that it collapses into any particular macrostate $X$ is given by the Born rule.\footnote{That is, $P( X) = \int_{X} |\psi(x)|^2 dx$. }

As such, quantum mechanics is not a candidate for a fundamental physical theory. It has two dynamical laws: the deterministic Schr\"odinger equation and the indeterministic collapse rule. What are the conditions for applying the former, and what are the conditions for applying the latter? Measurements and observations are extremely vague concepts. Take a concrete experimental apparatus for example. When should we treat it as part of the quantum system that evolves linearly and when should we treat it as an ``observer,'' i.e. something that stands outside the quantum system and collapses the wave function? That is, in short, the quantum measurement problem.\footnote{See \cite{bell1990against} and \cite{sep-qt-issues} for  introductions to the quantum measurement problem.} 

Various solutions have been proposed to solve the measurement problem. Bohmian mechanics (BM) solves it by preserving the Schr\"odinger dynamics, adding particles to the ontology, and an additional guidance equation for the particles' motion. Ghirardi-Rimini-Weber (GRW) theories postulate a spontaneous collapse mechanism, making wave function collapses independent of the observers. Everettian quantum mechanics (EQM) simply removes the collapse rules from standard quantum mechanics and suggest that there are many (emergent) worlds, corresponding to the branches of the wave function, which are all real. 

Both BM and GRW use probabilistic postulates to account for the Born rule. BM postulates the Quantum Equilibrium Distribution, which says that the initial particle configuration is distributed by the Born rule (see \cite{durr1992quantum}). GRW postulates probabilistic modification of the Schr\"odinger equation by which the wave function spontaneously collapses at a fixed rate and the center of the  collapse is distributed randomly according to  (something close to) the Born rule. EQM, developed and defended by  \cite{wallace2012emergent},   does not introduce any objective probabilities (but seeks to derive them from decision-theoretic  axioms). On BM and GRW, SP will postulate a fundamentally different kind of probabilities from the quantum mechanical ones. It would be desirable if they can be unified. On EQM, the aspiration is to come up with a theory that explains the probabilistic phenomena in nature, for which the objective statistical mechanical probabilities of SP will be an obstacle.\footnote{There have been some proposals of how to unify the two, such as Albert (2000), Ch.7 and \cite{wallace2011logic}. They rely on additional plausible conjectures.}

Recent work in the foundations of quantum mechanics  suggests that just as we can add particles in Bohmian mechanics (BM), we can add additional ontologies to GRW and Everettian theories: GRW with a flashy ontology (GRWf), GRW with a mass-density ontology (GRWm), and Everettian theory with a mass-density ontology (Sm).\footnote{See \cite{allori2013primitive}. Sm was introduced in \cite{allori2010many}. These are also called \textit{quantum mechanics with primitive ontologies}.} Let us  call them \emph{quantum theories with additional ontologies}. Unlike Bohmian particles, these additional ontologies are not variables independent from the wave function.

The above quantum theories---BM, GRW, GRWm, GRWf, EQM, Sm---make plausible the view that I call \emph{Wave Function Realism}: the universal quantum state is (1) ontic and (2) completely represented by the universal wave function. This is in contrast to the epistemic views about the wave function that maintain that the quantum state, represented by a wave function, corresponds to only our epistemic uncertainties over the actual state of the world. 

In short, the quantum Mentaculus contains the quantum versions of PH and SP that support the claim that typical initial microstates will be entropic. Such an understanding can be supplemented with further interpretations about the meaning of the wave function. But the ``marriage'' between the Mentaculus and Humeanism is not perfect; as  we discuss in \S2 and \S6, it leads to \emph{prima facie} problems  that seem to cry out for a different approach:
\begin{itemize}
\item Non-separability and non-narratability problems  if we keep the quantum state in the mosaic;
\item Complexity problems if we move the quantum state from the mosaic to the best system; 
	\item Supervenient-kind  and  vagueness problems of PH;
	\item (Not a problem but still worth mentioning: dualism of statistical mechanical probabilities and quantum mechanical probabilities.)
\end{itemize}


\subsection{The Wentaculus}

In this section, I construct an alternative framework---the Wentaculus.  In \S4 and \S5, I show that the Wentaculus  solves the problems above \emph{and} contains additional theoretical virtues. Here I briefly explain two components of the framework: (1) Density Matrix Realism and (2) the Initial Projection Hypothesis. (For a more systematic presentation, see \cite{chen2018IPH}.) I will also explain how Humeanism  provides motivations for the Wentaculus. 

\subsubsection{Density Matrix Realism}

In \S3.1, we saw that the quantum Mentaculus assigns probabilities over wave functions. Now, there is a well-known method of encoding the probabilities in the quantum state itself. Instead of saying that the wave function lies inside some subspace $\mathscr{H}_{\nu}$ and that there is a uniform probability distribution over the wave functions in  $\mathscr{H}_{\nu}$, quantum theory provides a compact way of putting these two pieces of information together into one mathematical gadget---a density matrix. The probability distribution over wave functions can be represented by a density matrix $\hat{W}_{\nu}$.\footnote{More precisely, the density matrix is equal to an integral over wave functions inside the unit sphere of the subspace with respect to the uniform distribution given by the surface area measure:  
$\hat{W}_{\nu} = \int_{\mathscr{S}(\mathscr{H}_{\nu})} \mu(d\psi) \ket{\psi} \bra{\psi}.$ Here is a more intuitive way of understanding the construction procedure. Start from the subspace $\mathscr{H}_{\nu}$. It is compatible with many vectors representing different initial wave functions. Take an arbitrary vector $\ket{\psi}$. We can construct a \emph{projection operator} (projecting to $\ket{\psi}$) as $\ket{\psi}\bra{\psi}$. If we apply $\ket{\psi} \bra{\psi}$ to any other vector $\ket{\phi}$, it will first take the inner product $\braket{\psi | \phi}$ and output a scalar $c$, which measures ``how much'' of $\ket{\phi}$ overlaps with $\ket{\psi}$. Then it multiplies the scalar to $\ket{\psi}$, which yields $c\ket{\psi}$. Take all the vectors on $\mathscr{S}(\mathscr{H}_{\nu})$, the unit sphere in the subspace, construct the corresponding projection operators, and then take the ``weighted sum'' over the projection operators. Since there is a continuous infinity of objects to sum over, instead of using an infinite sum, we use an integral over them with respect to the  surface area measure on the unit sphere $\mu(d\psi)$. This construction produces a density matrix that represents the probability distribution over the initial wave function. 
}

We should not be misled by the language here. Even though we  talk about ``constructing a density matrix from a collection of wave functions,'' there is a more \emph{intrinsic} way of understanding the   density matrix  that is independent of the wave functions. A density matrix is a well-defined object in its own right in  Hilbert space. Instead of constructing it from wave functions, we can  think of the above-mentioned density matrix as a simple object defined on  subspace $\mathscr{H}_{\nu}$. The object contains no more and no less information than what is contained in the subspace itself. It is called the normalized projection, whose mathematical representation can be written as follows:
\begin{equation}\label{ID}
\hat{W}_{\nu} = \frac{\mathbb{I}_{\nu}}{dim \mathscr{H}_{\nu}},
\end{equation}
This is the \emph{normalized projection} onto the subspace $\mathscr{H}_{\nu}$. The normalization is achieved by dividing the subspace identity operator $\mathbb{I}_{\nu}$ by the dimension of the subspace $dim \mathscr{H}_{\nu}$. The identity operator is restricted to the subspace: it does nothing to vectors contained inside the subspace and projects into the subspace everything that is not completely contained within.\footnote{Since the diagonal entries of $\hat{W}_{\nu} $ add up to $1$, it is a density matrix.} 

Given the intrinsic understanding of density matrices in Hilbert space, is there a sense we can provide an intrinsic understanding of it on configuration space that is independent from wave functions and equally objective? The answer is yes. We call this perspective Density Matrix Realism, in contrast to Wave Function Realism. Just as we can think of the wave function as a function that assigns values to the configuration space $\mathbb{R}^{3N}$, we can think of the density matrix as a function that assigns values to the Cartesian product of the configuration space with itself. Moreover, we can also think of it as a function that assigns values to every ordered pair of points in configuration space. 
 
Wave Function Realism is motivated by the idea that the wave function is central to the dynamics and the kinematics of quantum mechanics. In order to motivate Density Matrix Realism, we need to reformulate quantum mechanics directly in terms of a fundamental density matrix. This can be done in the following way.\footnote{Density Matrix Realism has already been suggested but not necessarily endorsed by some in the literature. For some recent examples, see \cite{durr2005role, maroney2005density},  \cite{wallace2010quantum} and \cite{wallace2011logic, wallace2012emergent}. What is new in this paper is the natural combination of Density Matrix Realism with PH in forming the Initial Projection Hypothesis (\S 2.2.2) and the argument for the Humean Unification based on that. }

First, the density matrix has an evolution equation analogous to that of the wave function. While the wave function obeys the Schr\"odinger equation, the density matrix obeys its generalization to mixed states---the von Neumann equation: 
\begin{equation}\label{VNM}
i \hbar \frac{d \hat{W}(t)}{d t} = [\hat{H},  \hat{W}],
\end{equation}
where the commutator bracket represents the linear evolution analogous to the linear evolution described by the Schr\"odinger equation. 

Second,  the Born rule distribution can be written in terms of the density matrix:
\begin{equation}\label{WBorn}
P(q)dq =  W(q, q) dq
\end{equation}

Third, we can reformulate BM, GRW, and EQM in terms of the density matrix.\footnote{For W-EQM, equation (\ref{VNM}) is all there is to govern the fundamental quantum state $W$. 

For W-EQM with a mass-density ontology, we can define the mass density function in terms of the density matrix:
\begin{equation}\label{mxt}
m(x,t) = \text{tr} (M(x) W(t)),
\end{equation}
where $M(x) = \sum_i m_i \delta (Q_i - x)$ is the mass-density operator, which is defined via the position operator $Q_i \psi (q_1, q_2, ... q_n)= q_i \psi (q_1, q_2, ... q_n) $. This allows us to determine the mass-density ontology at time $t$  via $W(t)$. 

For W-BM, we can postulate the guidance equation as follows:
\begin{equation}\label{WGE}
\frac{dQ_i}{dt} = \frac{\hbar}{m_i} \text{Im} \frac{\nabla_{q_{i}}  W (q, q', t)}{ W (q, q', t)} (q=q'=Q),
\end{equation}
 Finally, we can  impose a boundary condition similar to that of the Quantum Equilibrium Distribution: 
\begin{equation} \label{WQEH}
P(Q(t_0) \in dq) =  W (q, q, t_0) dq.
\end{equation}
Since the system is also equivariant, if the probability distribution holds at $t_0$, it holds at all times. Equivariance holds  because of the following continuity equation: 
$$\frac{\partial  W(q,q,t) }{\partial t} = -\text{div} ( W(q, q, t) v),$$ where $v$ denotes the velocity field generated via (\ref{WGE}.) This theory is first described in \cite{durr2005role} and \cite{maroney2005density}. \cite{durr2005role}  call this theory W-Bohmian mechanics. 

For W-GRW (first suggested in \cite{allori2013predictions}), between collapses, the density matrix will evolve unitarily according to the von Neumann equation. It collapses randomly, where the random time for an $N$-particle system is distributed with rate $N\lambda$, where $\lambda$ is of order $10^{-15}$ s$^{-1}$. At a random time when a collapse occur at ``particle'' $k$ at time $T^-$, the post-collapse density matrix at time $T^+$ is the following:
\begin{equation}\label{collapse}
W_{T^+} = \frac{\Lambda_k (X)^{1/2} W_{T^-} \Lambda_k (X)^{1/2}}{\text{tr} (W_{T^-} \Lambda_k (X)) },
\end{equation}
with $X$ distributed by the following probability density:
\begin{equation}\label{center}
\rho(x) = \text{tr} (W_{T^-} \Lambda_k (x)), 
\end{equation}
where $W_{T^+}$ is the post-collapse density matrix, $W_{T^-}$ is the pre-collapse density matrix, $X$ is the center of the actual collapse, and $\Lambda_k (x)$ is the collapse rate operator defined as follows:
$$\Lambda_k (x) = \frac{1}{(2\pi \sigma^2)^{3/2}} e^{-\frac{(Q_k -x)^2}{2\sigma^2}},$$ where $Q_k$ is the position operator of ``particle'' $k$, and $\sigma$ is a new constant of nature of order $10^{-7}$ m postulated in current GRW theories. 

For W-GRWm, we can let the density matrix determine the mass density function on space-time by (\ref{mxt}). For W-GRWf, we postulate  flashes that are the space-time events at the centers ($X$) of the W-GRW collapses. }
We can show that each reformulation of the realist quantum theory in terms of a universal density matrix $W$ is empirically equivalent to its wave-function counterpart, if on the latter theories the uncertainty over the universal wave function is represented by a statistical density matrix $W$.\footnote{This is because the predictions of quantum theory are probabilistic; it does not matter whether the density matrix we use to extract predictions is statistical or fundamental. See \cite{durr2005role, wallace2016probability}, and \cite{chen2019quantum1} for more detailed arguments. }  Therefore, these are also empirically adequate quantum theories without observers. We call these theories \emph{W-Bohmian mechanics, W-GRW theory}, and \emph{W-Everettian quantum mechanics}. Thus, we can think of $W$ as the central dynamical object in quantum mechanics that produces quantum mechanical phenomena and determine the behaviors of ``local beables.'' This makes possible an ``ontic'' interpretation of the density matrix: it is the complete description of the quantum state of the world; there is no more fundamental fact about which wave function is the actual one. Thus, the framework of Density Matrix Realism is a viable alternative to Wave Function Realism. 

Density Matrix Realism goes against the orthodox view, taught in many physics textbooks, about the role of density matrices in quantum mechanics. On the orthodox view, a suitably isolated system (such as a system that is not entangled with the environment or the universe as a whole) should be described as a pure state, while a mixed state corresponds to the epistemic uncertainty of the underlying pure state. But that view is rejected here. Given the viability of Density Matrix Realism,  and the possibility of describing microscopic motion using fundamental density matrices, we should not insist on describing suitably isolated systems using pure states only. Some systems may be in pure states but other systems (including the universe) can  be in fundamental mixed states. For example, the universe may be in a fundamental mixed state.  If so,  the universal quantum state is not pure; it does not correspond to a universal wave function.   The appropriate representation is given by a (mixed) universal density matrix that does not arise from an underlying pure state.  In the next section, I propose a natural choice of  the universal density matrix. Given the empirical equivalence between the orthodox theory and the density-matrix-realist counterpart, there are no experiments or observations we can perform to tell which one is true. We have to use super-empirical virtues to argue for one or the other.

\subsubsection{The Initial Projection Hypothesis}


 In the framework of Density Matrix Realism, the universal quantum state may be in a fundamental mixed state (although it does not have to be, as it can also be in fundamental pure states). Hence, we can consider reformulating the  low-entropy boundary condition as the constraint on possible initial density matrices (which can be either pure or mixed). However, just as there are many wave functions compatible with $\mathscr{H}_{PH}$, there are many  density matrices compatible with $\mathscr{H}_{PH}$, the PH subspace in the total Hilbert space.   One could construct a uniform probability distribution over the possible initial density matrices \citep{chen2020uniform}. 

However, that is not the only possibility. In fact,  Density Matrix Realism provides a much simpler constraint that combines PH and SP that is unavailable in the wave function framework. That is the  crucial insight here. I postulate that the initial density matrix of the universe is the simplest and most natural one associated with $\mathscr{H}_{PH}$: its normalized projection. It can be expressed as follows:
\begin{equation}\label{PHID}
\hat{W}_{IPH} (t_0) = \frac{\mathbb{I}_{PH}}{dim \mathscr{H}_{PH}},
\end{equation}
It is the identity operator on $\mathscr{H}_{PH}$ divided by the dimension of $\mathscr{H}_{PH}$. I label its Hilbert space representation as $\hat{W}_{IPH} (t_0)$. In the position representation, it is $W_{IPH} (q, q', t_0)$.  

This constraint is motivated by Humeanism. The goal of a Humean theorist is to come up with the simplest and most informative summary of the history of the world. If we can avoid the postulation of a probability distribution by making the initial state unique, then the Humean theorist would be motivated to do so. As we shall see in the next section, the postulate leads to further benefits from a Humean perspective. In contrast, there is no obvious candidate for the simplest or most natural wave function compatible with PH. 

Therefore, I propose that we add the following postulate to any quantum theory in the framework of Density Matrix Realism: 
\begin{description}
\item[Initial Projection Hypothesis:] The initial quantum state of the universe is $\hat{W}_{IPH} (t_0)$.
\end{description}
The Initial Projection Hypothesis (IPH) plays a similar role as that of PH. They both rule out many available initial states on the state space to   explain the time asymmetry in our universe. They carry the same information about initial entropy. PH selects the initial wave function to be one of the wave functions inside $\mathscr{H}_{PH}$, and IPH selects the initial density matrix to be the (unique) normalized projection on $\mathscr{H}_{PH}$. Both have exactly the same amount of entropy---that of $\mathscr{H}_{PH}$.\footnote{$S_B (\Psi_{PH}(t_0)) = S_B(W_{IPH} (t_0) ) = k_B \text{log} (\text{dim} \mathscr{H}_{PH} )$, 
where $S_B$ is the Boltzmann entropy function, $k_B$ is the Boltzmann constant, and ``dim'' counts the dimensionality of the subspace. }

However, there are crucial differences between IPH and PH. First, IPH picks out a unique initial quantum state of the universe while PH does not. In so far as the PH subspace can be unambiguously specified given some coarse-graining variables, the normalized projection can be unambiguously specified, and IPH also unambiguously specifies the initial state as $\hat{W}_{IPH} (t_0)$.\footnote{There are additional subtleties  about vagueness, which I explore in \S5.2. } In contrast, PH narrows down the initial wave function to the subspace  $\mathscr{H}_{PH}$, which is still compatible with an infinite number of different wave functions. 

Second,  IPH requires no further statistical mechanical probability distribution while PH needs to be supplemented with SP (or some measure of typicality). Since IPH chooses a \emph{unique} initial state, there is no need to add a probability weighting on the initial states compatible with IPH. In contrast,  PH is compatible with many wave functions, some of which will evolve to lower-entropy states.  Hence, PH needs to be supplemented with a statistical mechanical probability distribution (SP) that assigns high weight to the collection of ``good'' wave functions and low weight to the collection of ``bad'' ones. 

Because of these features of IPH, it seems that fundamental density matrices and the low-entropy initial condition are tailor-made for each other. That is the crucial insight we need to solve the problems on Humeanism. When we add IPH to Density Matrix Realism, we arrive at an alternative account of time's arrow in a quantum universe:
\begin{tcolorbox}
\centerline{\textbf{The Wentaculus}}
\begin{enumerate}
\item \textbf{Fundamental Dynamical Laws (FDL):} the quantum state of the universe is represented by a density matrix  $\hat{W}(t)$ that obeys the von Neumann equation (\ref{VNM}).\footnote{For GRW-type theories, the density matrix obeys the stochastic modification of the von Neumann equation described in footnote \#22. }
\item \textbf{The Initial Projection Hypothesis (IPH)}: at a temporal boundary of the universe, the density matrix is the normalized projection onto $\mathscr{H}_{PH}$,  a low-dimensional subspace of the total Hilbert space. (That is, the initial quantum state of the universe is $\hat{W}_{IPH} (t_0)$ as described in equation (\ref{PHID}).)
\end{enumerate}
\end{tcolorbox}

This is the W-version of the Mentaculus. Let us call it the \emph{Wentaculus}. To solve the quantum measurement problem, we can construct Bohmian, Everettian, and GRW versions of the Wentaculus.\footnote{The Wentaculus as it is will be sufficient for W$_{IPH}$-EQM. See \S4.2 for the the Bohmian version. 
}
Let us call these theories W$_{IPH}$-quantum theories. In \S4, I show that the Wentaculus naturally leads to a Humean unification of the origins of quantum entanglement and time asymmetry. The theoretical payoffs of the Humean unification are discussed in \S5. 

\section{The Humean Unification}

The density matrix formalism opens up a new possibility for understanding a time-asymmetric quantum-mechanical world: it can be described by the W$_{IPH}$-quantum theories of the Wentaculus. In this section, I show that Humeanism allows us to further simplify the theoretical structure, by unifying the sources of time asymmetry and quantum entanglement and removing the quantum state from the mosaic. First, I  argue for the Nomological Thesis. Second, I  show that Humeanism allows us to obtain a unified explanation of time asymmetry and quantum entanglement. Third, I discuss two worries one might have about the strategy. 

\subsection{The Nomological Thesis}

The classical Mentaculus (\S2) consist in three postulates---the fundamental dynamical equations, PH, and SP---all of which can be admitted, by the best-system account, as Humean laws. PH and SP are not the usual dynamical laws. In particular, PH is regarded as a Humean law even though it may look like just another contingent initial condition. Even before getting into Humeanism, there are pre-theoretic reasons that support its elite status as a fundamental physical law \citep{chen2020harvard}. For example, plausibly it plays a starring role in deriving the Second Law of Thermodynamics; and perhaps also in deriving the counterfactual asymmetries, the records asymmetry, the epistemic asymmetry, and influence asymmetry in time. They support the idea that PH is nomologically necessary and not merely contingent. Absent any further nomological explanations, PH is a fundamental law of nature and not a contingent initial condition. 

The Humean best-system account provides another argument for the nomological status of PH. Take for example the quantum Mentaculus. If we subtract PH and SP from the Mentaculus, the theory is much weaker. Let us call it the quantum Mentaculus$^-$:
\begin{tcolorbox}
\centerline{\textbf{The Quantum Mentaculus$^-$}}
\begin{description}
\item \textbf{Fundamental Dynamical Laws (FDL):} the quantum microstate of the universe is represented by a wave function $\Psi$ that obeys the Schr\"odinger equation $ i\hbar \frac{\partial \psi}{\partial t} = \hat{H} \psi$, where $\hat{H}$ encodes the fundamental interactions in quantum mechanics. 
\end{description}
\end{tcolorbox}
Since  Quantum Mentaculus$^-$ is time symmetric, it does not ground lawful generalizations such as the Second Law of Thermodynamics and many other temporal asymmetries. As such, it is much less informative than the Mentaculus. The Mentaculus$^-$ would be much more informative if we add to it PH and SP, and the cost in complexity is outweighed by the increase in strength. Hence, the Mentaculus is a better system than the Mentaculus$^-$. 

Moreover, PH and SP are not very complex. The uniform surface area measure, specified by SP, is a simple probability measure on the subspace. PH is simple in the macro-language specified in terms of the macro-variables such as temperature, volume, densities, and entropy.  In fact, PH could also be simple in the micro-language. A version of PH in the general relativistic cosmological context is the Weyl Curvature Hypothesis (WCH), which is a simple postulate about the initial geometry.\footnote{ In the context of discussing the origin of the Second Law of Thermodynamics in a   universe started with high homogeneity and isotropy, and the relationship between space-time geometry and entropy, Penrose proposes a hypothesis: 
\begin{quote}
I propose, then, that there should be complete lack of chaos in the initial \emph{geometry}. We need, in any case, some kind of low-entropy constraint on the initial state. But thermal equilibrium apparently held (at least very closely so) for the \emph{matter} (including radiation) in the early stages. So the `lowness' of the initial entropy was not a result of some special matter distribution, but, instead, of some very special initial spacetime geometry. The indications of [previous sections], in particular, are that this restriction on the early geometry should be something like: \emph{the Weyl curvature $C_{abcd}$ vanishes at any initial singularity}. (\cite{penrose1979singularities}, p.630, emphasis original)
\end{quote}
The Weyl curvature tensor $C_{abcd}$ is the traceless part of the Riemann curvature tensor $R_{abcd}$. It is not fixed completely by the stress-energy tensor and thus has independent degrees of freedom in Einstein's general theory of relativity. Since the entropy of matter distribution is quite high, the origin of thermodynamic asymmetry should be due to the low entropy in geometry, which corresponds very roughly to the vanishing of the Weyl curvature tensor. The Weyl Curvature Hypothesis is simple to state in the language of general relativity. } To have a complete quantum generalization of WCH would require a theory of quantum gravity, which is still work in progress. However, there are reasons to be hopeful. For example, the generalization of WCH to  Loop Quantum Cosmology program yields the Quantum Homogeneity and Isotropy Hypothesis (QHIH) according to which the initial quantum state has to come from a small subset of possible states with low entropy. It retains the general features of WCH but also introduces some vagueness (about the proper duration of the Planck regime), which is to be expected for any hypothesis that relies on some kind of coarse-graining.\footnote{See \cite{ashtekar2016initial} and \cite{ashtekar2016quantum}. } 

We have good reasons to think that the quantum Mentaculus could be the best system. Thus, we have good reasons to think that PH and SP are parts of the best system. On the modified Humean theory of laws and objective probabilities, it follows that PH is a Humean law of nature and SP specifies objective probabilities in the world. 

Similarly, the best system from the point of view of Density Matrix Realism is the Wentaculus. Given the crucial role  IPH plays in the Wentaculus, it is plausible that  IPH should be regarded as a Humean law if the Wentaculus is the best system. After all,  IPH has the same informational content as PH. They both specify a low-entropy initial condition. Moreover, IPH is as simple as PH+SP. They pick out the same density matrix in the low-entropy subspace. 

Hence, in so far as we have good reasons to take PH to be a Humean law  if the quantum Mentaculus is true, we have equally good reasons to take IPH to be a Humean law if the Wentaculus is true. That is, if Wentaculus is the right theory of the actual world, then we have good reasons to confer  (Humean) nomological status to IPH that are on a par with our reasons for conferring (Humean) nomological status to PH. But how do we know which is true: the Mentaculus or the Wentaculus? Here we encounter a case of underdetermination by evidence. The two theories are empirically equivalent: no amount of empirical evidence can settle the question which one is correct. Nevertheless, we can consider the super-empirical virtues, some of which will be discussed in \S5. 

Both PH and IPH are Humean laws  about the initial quantum state. As discussed before, it is controversial what the quantum states represent. But what is the nature of the initial quantum state? A promising answer suggests that it is  nomological. 
\begin{description}
\item[The Nomological Thesis:] The initial quantum state of the world is nomological, i.e. it is on a par with laws of nature. 
\end{description}
The Nomological Thesis, on the one hand, is in tension with the complexity issue in the quantum Mentaculus. Even though PH is simple, the wave function compatible with PH is unlikely to be simple enough to be nomological. The Humeans could follow \cite{goldstein1996bohmian} and claim that the Wheeler-DeWitt equation in quantum gravity would produce a time-independent wave function that may also be simple enough. Even though the Wheeler-DeWitt equation leads to fascinating scientific and interpretational questions, the Humeans who endorse this strategy faces several challenges. First, a technical challenge: given the time-independence of the wave function, the time asymmetry cannot be described in terms of the entropy of the universal wave function, and that would require significant changes to the Mentaculus program. Moreover, it does not follow that a time-independent wave function will be guaranteed to be sufficiently simple. We need  additional reasons to support that conjecture. Second, a strategy question: do Humeans want to tie the tenability of Humeanism to a particular equation in  quantum gravity? It seems reasonable to seek alternative ways to defend the Nomological Thesis. If for nothing else, defenders of Humeanism would be wise to find multiple ways to reconcile their view with fundamental physics. 

On the other hand, the Wentaculus transforms the situation. Since IPH is a law, and since IPH completely specifies the initial quantum state, $W_{IPH} (t_0)$ is no more complex than IPH itself. So if IPH is simple enough to be a law, then $W_{IPH} (t_0)$ is simple enough to be nomological. (In contrast, even though PH is simple enough, the initial wave function in the Mentaculus may not be simple.) Hence, in the Wentaculus (but not in the Mentaculus), we can easily remove the complexity obstacle by regarding the initial quantum state to be on a par with laws of nature. 

I propose that the Humeans remove the initial quantum state from the mosaic and move it to the best system. Moving $W_{IPH} (t_0)$ to the best system will not overburden the best system with too much complexity. After all, IPH is already an axiom of  the Wentaculus best system and IPH completely specifies $W_{IPH} (t_0)$ . (In contrast, PH does not contain all the information to pin down the initial microstate of the quantum Mentaculus (wave function).)  However, after we remove the quantum states from the mosaic, we need to make sure the mosaic is not empty. To do so, we can postulate the local material ontology of  as BM, Sm, GRWm, or GRWf.  The fundamental ontology, in each of these theories, will be the particles, matter densities, or flashes, which are separable. 

What about later quantum states $W_{IPH} (t_1)$, $W_{IPH} (t_2)$,..., and so on? Do we need to postulate them in the mosaic? That is not necessary. For unitary quantum theories such as BM and Sm, their information can be directly derived from  the von Neumann equation, which is also in the best system. For stochastic theories, the initial quantum state (in the best system) can specify a complete probability distribution and conditional probabilities over possible mosaic histories.\footnote{ The GRW route is different from the way we think about making predictions on GRW theories.  That might mean that the GRW route of the Humean unification is less natural than the route on unitary theories. Hence, the Humean unification may be sensitive to empirical questions about whether GRW is correct, and whether quantum theory is exact. Even if one is bothered by this sensitivity, one need not give it too much weight all things considered. So far, all experimental tests to find violations of unitary dynamics and the deviations from exact Born rule have confirmed exact quantum theories such as BM and EQM; we have not found any confirmation of GRW over its rivals. For a review, see \cite{feldmann2012parameter}. }

\subsection{A Unified Explanation}

If the initial quantum state is nomological, then the Humean best system contains all of its information: without adding any contingent fact from the mosaic, we can deduce its state at any time from just the best system alone. We do not need to, as we standardly do, independently specify the initial condition of the quantum state as a contingent fact in the mosaic.  In the Wentaculus framework, we can remove the quantum state from the mosaic without losing any information about entanglement  correlations, for such information is already contained in the best system. And the mosaic will not be empty---it will still contain the local beables such as particles, matter densities, and flashes, which make up pointers, tables, and chairs. From a Humean point of view, the best system (now the Wentaculus plus the values of the initial quantum state) supervenes on the mosaic. 

Take W$_{IPH}$-BM for example. Let us write down the mosaic + best system package without the Humean unification: 
\begin{tcolorbox}

\begin{description}
\item[The W$_{IPH}$-BM mosaic:] particle trajectories $Q(t)$ on physical space-time and the quantum state $W_{IPH}(t)$.

\item[The W$_{IPH}$-BM best system:]  four equations---the simplest and strongest axioms summarizing the mosaic: 

\begin{itemize}
\item[(A)] The von Neumann equation: $i \hbar \frac{\partial \hat{W}}{\partial t} = [\hat{H},  \hat{W}],$

\item[(B)] The Initial Projection Hypothesis: $\hat{W}_{IPH} (t_0)  = \frac{I_{PH}}{dim \mathscr{H}_{PH}}$

\item[(C)] The W-Quantum Equilibrium Distribution: $P(Q(t_0) \in dq) =  W_{IPH} (q, q, t_0) dq,$

\item[(D)] The W-guidance equation: $\frac{dQ_i}{dt} = \frac{\hbar}{m_i} \text{Im} \frac{\nabla_{q_{i}}  W_{IPH} (q, q', t)}{ W_{IPH} (q, q', t)} (q=q'=Q).$
\end{itemize}
\end{description}
\end{tcolorbox}

What would it look like under the new proposal? It will have fewer things in the mosaic and fewer equations in the best system. 

 \begin{tcolorbox}

\begin{description}
\item[The W$_{IPH}$-BM mosaic:] particle trajectories $Q(t)$ on physical space-time.

\item[The W$_{IPH}$-BM best system:]  three equations---the simplest and strongest axioms summarizing the mosaic: 

\begin{itemize}

\item[(A)] The Initial Projection Hypothesis: $\hat{W}_{IPH} (t_0)  = \frac{I_{PH}}{dim \mathscr{H}_{PH}}$
\item[(B)] The W-Quantum Equilibrium Distribution: $P(Q(t_0) \in dq) =  W_{IPH} (q, q, t_0) dq,$
\item[(C)]  The combined equation: $\frac{dQ_i}{dt} = 
\frac{\hbar}{m_i} \text{Im} \frac{\nabla_{q_{i}}   \bra{q} e^{-i \hat{H} t/\hbar} \hat{W}_{IPH} ( t_0) e^{i \hat{H} t/\hbar} \ket{q'} }{ \bra{q} e^{-i \hat{H} t/\hbar} \hat{W}_{IPH} (t_0) e^{i \hat{H} t/\hbar} \ket{q'}} (q=q'=Q)$

\end{itemize}

\end{description}
\end{tcolorbox}
In this theory, the mosaic no longer contains the quantum state. IPH still postulates the values of the initial quantum state $\hat{W}_{IPH} (t_0)$. But it is dispensable. The role it plays in the best system above is to specify  the velocity field of particle configuration and the initial probability distribution. We can rewrite any occurrence of $\hat{W}_{IPH} (t_0)$ in terms of its explicit functional form. We can construct similar Humean interpretations of W$_{IPH}$-GRWm, W$_{IPH}$-GRWf, and W$_{IPH}$-Sm.

The nomological role of the quantum state here is similar to that of the Hamiltonian function in classical mechanics. The Hamiltonian specifies the interactions or the ``forces'' among the component systems. The Hamiltonian is on par with the classical laws of motion as it is a simple part of the Hamiltonian equations. That is, if we expand the Hamiltonian function as a function of the variables for things in the mosaic (positions and velocities of particles), the equation is still simple.\footnote{The Hamiltonian equations are:\begin{equation}
\frac{\partial \boldsymbol{q_i}}{\partial t} = \frac{\partial H}{\partial \boldsymbol{p_i}} \text{  ,  } \frac{\partial \boldsymbol{p_i}}{\partial t} = - \frac{\partial H}{\partial \boldsymbol{q_i}}.
\end{equation}
The Hamiltonian function is specified as follows:
\begin{equation}
H(\boldsymbol{q}, \boldsymbol{p}) = \sum_{i=1}^{N} \frac{p_i^2}{2m_i} + \mathop{\sum\sum}_{1 \leq k < l \leq N} V_{k,l} (|q_k - q_l|),
\end{equation}
where $V_{k,l}$ is a simple formula for the pair-wise interactions. } Similarly, the quantum state is a simple part of the von Neumann equation. However, an important difference is that equation (C) is time-dependent, while the Hamiltonian equations of motion are time-independent.

I call such an interpretive strategy the \emph{Humean unification}. 
 On this proposal, we accept the Wentaculus, remove the quantum state from the Humean mosaic, and use a Humean law (IPH) to specify both the quantum data and the low-entropy initial condition. How, then, does one  explain the phenomena of quantum entanglement? How can systems at space-like separation be perfectly correlated, if there is no fact about quantum entanglement in the mosaic?  The Humean unification provides purely nomic explanations. There is a law that specifies the quantum entanglement of all the systems at $t_0$, from which we can use the von Neumann equation to derive quantum entanglement at a later time $t$. The equation for local beables (such as (\ref{mxt}) and (\ref{WGE})) will then determine the behavior of objects in space-time: e.g. if Alice were to  observe ``Spin Up'' then Bob would observe ``Spin Down.'' (Both the observers and the observed systems will be made out of the local beables and not of the quantum state.) Moreover, the law that specifies the quantum initial condition---IPH---is the same law that specifies the low-entropy initial condition. Hence, IPH is the origin of both quantum entanglement and time asymmetry.

The Humean unification provides a unified view on the sources of quantum entanglement and time asymmetry. In the Mentaculus, they have distinct sources---one from the macroscopic Past Hypothesis and the other from a microscopic wave function. But in the Wentaculus, it is one and the same density matrix---$\hat{W}_{IPH} (t_0)$.

The time-dependence of the equation of motion (such as the combined equation (C)) in the unified best system further suggests that there is intertwining between the two. In the Mentaculus, the theory as a whole is not time-translation-invariant, because PH applies only at a particular time. However, we can still understand the sense in which the Mentaculus is still time-translation-invariant: we can separate the dynamics from the initial condition; the dynamics is invariant even though the lawful initial condition is not. But in the Wentaculus, after the Humean unification,  there is no such clean separation. The two, as it were, are genuinely unified into one law, so the theory is not fundamentally time-translationally invariant. 

The violation of time-translation invariance should be viewed not as a bug but  a feature of the Humean unification. After all, is it reasonable to insist on saving the symmetry at the expense of everything else? Of course not. Does this particular violation make the theory more complicated? No; in fact the theory becomes simpler because of it. Symmetries are only \textit{defeasible} indicators for simplicity, and in this case we have clear reasons to think that time-translation invariance does not point to the simpler one. 

Can we still make sense of  time-translation invariance somehow? Yes.  At the emergent level of effective dynamics and subsystem analysis, we can still retain a time-translation-invariant non-fundamental dynamics. For many subsystems, they will have (non-fundamental) subsystem density matrices that still obey time-translation-invariant (effective) laws. 

Even if one is worried about the loss of (fundamental) time-translation invariance, the theoretical cost should be viewed in the context of and in balance with  the numerous theoretical advantages (\S5).  The balance of considerations on the whole favors understanding the Wentaculus through the Humean unification.

\subsection{New Wrinkles}

The Humean unification  leads to some new wrinkles. I focus on two here: the complexity issue and the classical maneuver. 

(1) The complexity worry. The Initial Projection Hypothesis selects a density matrix that is mathematically equivalent to (though metaphysically prior than) a ``disjunction'' of wave functions with a uniform probability distribution over them. If any wave function in the disjunct is overly complex, shouldn't the whole disjunction, and therefore the initial density matrix, be overly complex? How is that compatible with the earlier claim that the Initial Projection Hypothesis as well as the initial density matrix are simple enough to be nomological? The intuition behind this worry is that the disjunction inherits whatever complexity that is in the disjuncts, and the initial density matrix $\hat{W}_{IPH} (t_0)$ will be highly complicated, and perhaps even more so than a typical wave function. So our criticism that the standard wave function nomological view introduces overly complex laws will apply to the Humean unification strategy. 

That is incorrect. I offer a counterexample to the intuition and a positive argument for the simplicity of the initial density matrix. Let us consider classical mechanics governed by $F=ma$. We can think of the content of $F=ma$ as given by the disjunction of all the solutions to that equation, namely the disjunction of all complete trajectories of any number of point particles that classical mechanics allows. Most of those trajectories will be highly complex. However, $F=ma$ is a simple law, even though it is informationally equivalent to the complete disjunction of its possible solutions.  Positive argument: there are ways of understanding the density matrix that is independent of the collection of wave functions; the quantum state space (Hilbert space) permits a straightforward, intrinsic, and geometrical understanding of the initial density matrix selected by IPH. In fact, it is  (modulo the normalization constant)  equivalent to the subspace itself. While an individual vector in the subspace (the wave function) may require many coordinate numbers to pick out, the subspace requires much less information to specify.  The availability of the intrinsic understanding of the fundamental density matrix $\hat{W}_{IPH} (t_0)$ also responds to the supervenient-kind problem raised in \S2.

(2) The classical maneuver. 
One might  worry that the same ``trick'' can be played in the classical context. This means that the Humean unification is too easy to achieve and therefore trivial. On first glance, the suggested maneuver is to take the ``probability distribution'' ($\rho$)  as ``ontic'' or ``nomic.'' The same thing can presumably be done in the classical context (see \cite{McCoySMS} for example), where the probability distribution on phase space can be given a similarly ontic or nomic interpretation, thus avoiding the problems in the classical domain as well. If that is possible, moreover, it seems to show that either we have proven too much, or that it does not depend on the details of quantum theory. 

However,  that is a mistake. First, it is much less natural to give an ontic or nomic interpretation of the  probability distribution in classical statistical mechanics. If we use the same idea in the classical domain, we will likely get a many-worlds version of classical mechanics or lose determinism. The classical probability distribution $\rho$ plays no dynamical role (unlike the density matrix in the W-quantum theories). Moreover, since $\rho$ follows the Hamiltonian dynamics, it will in general be supported on many macroscopically distinct regions on phase space. If we reify $\rho$ as ontic and do not modify the dynamics, then we arrive at a many-worlds theory. If we modify the dynamics to introduce objective ``collapses'' of $\rho$ that take it to some ``branch'' of $\rho$, it will look much more artificial and complex than the original deterministic Hamiltonian theory. In contrast, on each of the three interpretations of QM, the artificial effects do not arise on the Wentaculus. The Bohmian version remains deterministic (and single-world), the GRW version remains stochastic (and single-world), and the Everettian / many-worlds version is still deterministic.  On the other hand, even if a classical extension of our maneuver is possible, it is unclear how it makes the quantum case trivial, since presumably both require different choices of the ontology and the dynamics.

\section{Fruits of the Humean Unification}

The Humean unification strategy has implications for what the Humean mosaic and the best system look like. In this section, I list some advantages of the new proposal. 

\subsection{The Mosaic: Simpler, Separable, and Narratable}

Under the Humean unification, the  mosaic becomes simpler, separable, and  narratable. In standard quantum theories, it is plausible that the quantum state represents something in the mosaic. This leads to a non-separable entity or relation that violates one of the tenets of Humean supervenience: that the mosaic must be separable. The Humean unification lifts the quantum state from the mosaic into the best system (without adding too much complexity to the best system). It simplifies the mosaic by removing the quantum state ontology and postulating local matters in spacetime---particles, matter density, or flashes. The  separability of the mosaic is now restored. 

This result provides a response  to \cite{MaudlinMWP}'s influential argument that Humean supervenience is incompatible with quantum mechanics. It does so without incurring new costs (see contrast with rival proposals in \S6) and with many new benefits. 

An under-appreciated consequence of this result is that it also responds to \cite{albert2015after}'s recent  argument that quantum entanglement is incompatible with full Lorentz invariance.\footnote{I am indebted to discussions with Sheldon Goldstein and Ezra Rubenstein on this point.} This is best seen in W$_{IPH}$-Sm, which aspires to be a fully Lorentz-invariant theory. Albert (2015) shows  that the conjunction of Lorentz invariance and entanglement is inconsistent with narratability, the idea that the full history of the world can be narrated in a single temporal sequence, and other ways of narrating it will be its geometrical transformations. To describe a narratable mosaic temporally, it suffices to list facts along one foliation (with the rest being geometrical transformations); to describe a non-narratable mosaic such as the one that contains entanglement relations, we need to list facts along every foliation, which is infinitely more complex. Thus, one can understand narratability as something akin to descriptive parsimony, which is a desirable but defeasible virtue to be balanced with other considerations.\footnote{After all, one can consistently insist on  specifying states only directly on the mosaic and not through any temporal sequences picked out along some foliation.} In so far as one finds narratability a plausible principle,  there is tension between Lorentz invariance and quantum entanglement (which is a purely kinematic notion).  

However, Albert's argument presupposes that there are facts about the entanglement relations in the mosaic. Now, if we remove the quantum state from the mosaic, the trouble-maker is gone, and the mosaic can be both Lorentz invariant and narratable. This avoids the narratability failure mentioned earlier. (A similar result holds for W$_{IPH}$-GRWm and W$_{IPH}$-GRWf.)  By allowing the mosaic to be fully narratable and allowing the laws to be fully Lorentz invariant, the Humean unification could lead to further simplification of the mosaic and the best system. By  removing the conflict between Lorentz invariance and narratability,  Humean unification removes some tension between quantum mechanics and special relativity.\footnote{To be sure, there is still the issue of quantum non-locality.} 

In summary, the Humean unification strategy simplifies the mosaic by removing the quantum entanglement relations, restoring the separability and narratability of the fundamental  ontology. It also has a better chance of reconciling quantum theory with full Lorentz invariance (for those theories that have such aspirations). 

\subsection{The Best System: Simpler, Less Vague, and More Unified}

The Humean unification also leads to some important modifications of the Humean best system, making it simpler, less vague, and more unified. By lifting the quantum state from the mosaic and placing it in the best system, \emph{prima facie} we increase the complexity of the best system by exactly the amount of complexity of the quantum state. If the quantum state is highly complex, then the resultant best system will be complex as well. Even if the mosaic becomes simpler and separable, the costs may still be too high. (That may be the case for some of our rivals. See \S6.) However, given the analysis in \S4 about the simplicity of PH and the IPH, the initial quantum state is simple enough to be nomological. Making $\hat{W}_{IPH} (t_0)$ on a par with Humean laws does not overburden the best system. In fact, since the Wentaculus best system already contains IPH, and IPH completely specifies the values of $\hat{W}_{IPH} (t_0)$, there is no added complexity when we lift $\hat{W}_{IPH} (t_0)$ to the best system.  That is a significant advantage over other versions of quantum Humeanism where the focus is on the quantum wave function that is typically much more complex than the initial density matrix we chose.  

The Humean unification simplifies the best system in another way: it eliminates fundamental statistical mechanical probabilities in the best system \citep{chen2018valia}. In the quantum Mentaculus, there are two kinds of probabilities: the quantum probabilities prescribed by the Born rule (or the quantum equilibrium distribution in BM, the collapse probabilities in GRW, and the non-physical decision-theoretic or \emph{de se}  probabilities in EQM) and the statistical mechanical probabilities prescribed by the Statistical Postulate (SP). It would be desirable to unify the two sources of probabilities in the theory. \cite{albert2000time}\S7 suggests a different strategy  in the GRW framework, relying on a plausible conjecture about GRW collapses and ``thermodynamically abnormal'' region in Hilbert space. \cite{wallace2011logic} proposes we replace the SP by a simplicity constraint, relying on a conjecture about simplicity and reversibility. On the Humean unification, however, we have a completely general way of getting rid of SP. By choosing a unique and natural initial density matrix, we no longer have an infinity of possible initial microstates. There is just one state to choose from and we no longer need any probability distribution over possible  initial microstates. The Humean unification provides a simple and general way to avoid the dualism of probabilities. This is achieved by making SP simply unnecessary.\footnote{For W$_{IPH}$-Everettian theories, since the Born rule is not supposed to be fundamental, the elimination of SP means that there are no objective probabilities in the world. The actual world is  nomologically necessary. The theory will become \emph{strongly deterministic} in the sense of \cite{roger1989emperor}.}

In contrast to the quantum Mentaculus (and the classical Mentaculus), the Humean unification contains less vagueness. PH and SP are exact only when we choose some arbitrary coarse-graining variables in nature. On the Mentaculus, the boundary of the  set of microstates corresponding to the low-entropy macrostate, after a certain level of precision, will be entirely arbitrary. There is nothing in nature that pins it down. Beyond a level of precision, the exact boundary of the subspace makes no difference to how things are behaving in physical space and what their probabilities are. The same is true for the Statistical Postulate: the support of the probability distribution becomes exact only after we impose some arbitrary choices, and the exact values do not matter after a certain level of precision. So it is best to think of PH and SP as vague postulates and vague Humean laws in the best system of the Mentaculus. The situation is  different under the Wentaculus picture. The exact values of the initial density matrix makes a difference to the world just as importantly as constants of nature---different values will typically lead to different microscopic behaviors and probabilities.  The exact values of the initial quantum state is by no means arbitrary---$\hat{W}_{IPH} (t_0)$ plays a central role in the fundamental micro-dynamics; it makes a difference to the exact Bohmian velocity field, GRW collapse probabilities, and configuration of matter densities, etc. That is why we can eliminate the vagueness of IPH without objectionable arbitrariness \citep{chen2018NV}. 

Another advantage of the Humean unification is that it leads to more unity in the best system. First, it provides mathematical unity between quantum mechanics and quantum statistical mechanics. Quantum statistical mechanics makes extensive use of density matrices. Quantum mechanics, on the other hand, has often been formulated in terms of  wave functions. From the perspective of the Humean unification, density matrix is the central object in both theories. A wave function only arises in special circumstances when the density matrix is pure. Second,  there is an increase in the dynamical unity in some theories. In BM with spin, there does not exist a conditional wave function since the particles have only positions but  no spin degrees of freedom. Hence, in general, for the subsystems in a W-BM universe, there is only a conditional density matrix instead of a conditional wave function.  As a result, the guidance equation for many subsystems  (that are suitably isolated from the environment) will be the W-guidance equation (\ref{WGE}) that refers to a conditional density matrix even in standard BM, while the guidance equation for the whole universe will be the usual guidance equation that refers to a wave function.\footnote{See \cite{durr2005role} for more details. The standard guidance equation in BM under a universal wave function is: \begin{equation}\label{GE}
 \frac{dQ_i}{dt} = \frac{\hbar}{m_i} \text{Im} \frac{\psi^* \nabla_i \psi }{\psi^*  \psi} 
\end{equation}} Therefore, the dynamics for the universe will be importantly different from the dynamics for the subsystems for standard BM. In contrast, in W$_{IPH}$-BM, the guidance equation for the whole universe and that for the subsystems (that are suitably isolated from the environment) will be the same---(\ref{WGE}). Whether dynamical unity holds for  W$_{IPH}$-GRWm and W$_{IPH}$-Sm will require a further analysis of the subsystem dynamics in those theories. But in any case, those theories also witness an increase in kinematic unity: both the universe and typical subsystems will be in mixed states described by density matrices. This is in contrast to the Mentaculus picture, where most subsystems are in mixed states (due to entanglement) while the universe is in a pure state.

\section{Comparisons}

In this section, I discuss two other versions of (wave-function) quantum Humeanism that are motivated by the problem of quantum entanglement but not the problem of time asymmetry.  The new version of (density-matrix) quantum Humeanism suggested above  has advantages over both. 

\subsection{Wave Function Humeanism in a High-Dimensional Space}

One of the earliest proposals of how to reconcile  Humeanism with quantum entanglement is that of  \cite{LoewerHS}. Loewer argues that if we adopt David Albert's high-dimensional space fundamentalism, the idea that the fundamental space of the world is the ``configuration space,'' then the quantum state (represented by a wave function) is entirely separable in that fundamental space. We can reify that as the Humean mosaic. Entanglement and non-locality are merely manifestations of our perception in the low-dimensional space, which is not fundamental. 

The move from a low-dimensional fundamental space to a high-dimensional one is a radical move \citep{emery2017against}. It is  revisionary from the Humean perspective. On Lewis's original formulation, the Humean mosaic are facts about the physical space-time (or some low-dimensional manifold). But it is also revisionary from the ordinary scientific perspective. There are important reasons to take something like the physical space-time to be fundamental, as it underlies many important symmetries in physics, including Lorentz invariance. They will be difficult to recover from the high-dimensional point of view.\footnote{See \cite{allori2013primitive} for related reasons.  \cite{ChenOurFund} argues that the low-dimensional view provides better explanation of the Symmetrization Postulate than the high-dimensional view does. \cite{emery2017against} provides reasons to think that other things being equal we should prefer the more common-sense view of  physical reality.}  In contrast,  the Humean unification does not require such radical revisions. 

\subsection{Wave Function Humeanism in a Low-Dimensional Space}

The second class of proposals of reconciling Humeanism with quantum entanglement is  discussed in \cite{miller2013quantum, esfeld2014quantum, bhogal2015humean, callender2015one},  \cite{esfeld2017minimalist}. On this proposal, the quantum state of the universe represented by a wave function is part of the Humean best system. There are three ways to interpret this proposal. 

First interpretation: the wave function itself represents a simple and informative Humean law of nature (like the classical Hamiltonian). Miller and Callender are close to endorse this view. However, it requires that the universal wave function to be extremely simple, or at least simpler than the complete facts about local beables through all time. Otherwise the  system containing the wave function would not win the competition for being the best system. It faces the \emph{prima facie} problem that the universal wave function may not be simple enough to be nomological.
 
Second interpretation:   the wave function supervenes on the mosaic and it participates in the dynamical laws of nature. On this view, the Humean laws refer to the wave function, but the wave function itself is not ``nomological'' in the usual sense. Instead, the wave function represents a physical state---albeit a non-fundamental one. As such, both the wave function and the dynamical laws supervene on the mosaic. This interpretation is closest to the proposal of Bhogal and Perry.  The natural question arises again: does the wave function have to be  relatively simple? They seem to think so and suggest that the universal wave function (together with the dynamical laws) will be simpler than the mosaic itself. However, this is far from clear: even when the wave function helps systematize the mosaic (as their examples suggest), it is no trivial task to establish the simplicity of the universal wave function (relative to the mosaic).  If, on the other hand, we need to relax our criterion of simplicity, we may ask whether it is too ``liberal.'' Perhaps there is a system that contains the simplest wave function over all rivals; still in that case the wave function could be, given the standard scientific criterion of simplicity,   too complex to be on par with laws. That would count as a (defeasible) consideration against the interpretation. The worry is dissolved if none of the alternatives is better. However, as we argued in \S4 and \S5, there exists another system (which is compatible with the evidence we have) that is simpler than this one. 

Third interpretation: the wave function is merely a variable that we introduce into the best system to describe the mosaic. \cite{esfeld2017minimalist} hold this view. On their proposal, only point particles are fundamental, and we can interpret every other bit of the theory as in the best system. It is an open question whether there are any principled constraints on the proposal. If there are no principled constraints, then (in classical physics) we could follow their procedure and regard the electromagnetic fields as part of the best system. But the electromagnetic fields are usually highly complicated. Given the standard view that the electromagnetic fields are part of the material ontology, tying Humeanism to the radical view (that allows us to Humeanize the electromagnetic field) leads to undesirable consequences. Moreover,  it seems much more instrumentalist than the original Humean proposal, which \emph{aspires} to be realist. Humeanism requires delicate balances between objectivism and pragmatism, but at least the original Humean proposal has principled constraints on what goes into the best system and what goes into the mosaic, and such constraints are supposed to be compatible with scientific practice. 

In short, there are \emph{prima facie} problems facing  these versions of wave function Humeanism.  In contrast,  the Humean unification avoids these problems, as we know that the initial quantum state is simple and unique, and the proposal is fully realist (or as realist as Humeanism about laws can be). Moreover, the Humean Unification has many other theoretical virtues (\S5). 
  
Nevertheless, the Humean unification strategy is not incompatible with  the views discussed above. In fact, there is much common ground; our proposal can be seen as friendly extension to some of the views above. For example, towards the end of her insightful essay, \cite{miller2013quantum} is rightly worried about the potential slippery slope in the strategy she proposes, and she suggests we   find  principled limits to curb over-Humeanization or  the``narcissism'' if one  tries to ``Humeanize'' away everything one does not like in the ontology, including all particles outside one's brain.  Our framework can be seen as implicitly suggesting such a limit: carry out Humean unification \emph{only when} you can show that the best system will not be over-burdened by extra complexity \emph{and} that there will be additional fruits of unification beyond solving the original problems. Moreover, our framework can be combined with \cite{bhogal2015humean}'s proposal to simplify the supervenient L-state, and with \cite{esfeld2017minimalist}'s proposal to further simplify their ``super-Humean'' best-system. I leave that to future work.

\section{Conclusion}
The origin of time asymmetry in our universe lies in its boundary condition---a low-entropy state now called the \emph{Past Hypothesis}. In this paper, I replace it with the Initial Projection Hypothesis (IPH) and construct a new class of density-matrix-realist quantum theories---W$_{IPH}$-theories, or the Wentaculus. They allow  Humeans to use the best system to specify a simple and unique initial quantum state.  This provides a ``Humean unification'' of the explanations for quantum entanglement and time asymmetry. The data in IPH, with the help of  the density-matrix dynamics, gives rise to both time-asymmetric and quantum phenomena. The resultant theory has a separable and narratable mosaic as well as a simpler, less vague, and more unified law system. It turns out that a variety of problems become easier to solve on the new theory. 
 
Can a non-Humean appreciate the new theory? I think so. 
On the Mentaculus, PH is a macroscopic constraint that does not directly play a role in the microdynamics.  In contrast, IPH in the Wentaculus specifies an initial density matrix that is both the initial macrostate and the initial microstate.  The  low-entropy constraint gets connected to the microdynamics.  The initial density matrix directly determines the Bohmian velocity field, the GRW collapse probabilities, and the Everettian branching structure. Hence, it plays a role analogous to the classical Hamiltonian function, and as such it fits the  non-Humean ``governing'' conception of laws. 

In this sense, we can develop  a non-Humean counterpart to the Humean unification strategy. Impressed by the success of the Humean project, a non-Humean can accept the same mathematical theory but give it a different metaphysical interpretation. While the Humean understand the axioms of the best system as mere summaries of the mosaic, a non-Humean can understand those axioms as fundamental laws that govern the mosaic. On Humeanism, the theoretical virtues of simplicity, informativeness, and fit may be constitutive of what laws are, but they can be considered as epistemic guides for finding  the fundamental (non-supervenient) laws on non-Humeanism.\footnote{Can a quantum state monist appreciate the new theory? I think so. She can at least appreciate the W$_{IPH}$-quantum theories without local beables: W$_{IPH}$-GRW and W$_{IPH}$-EQM. The crucial step in the Humean unification---\S4---would not be possible. The quantum state will have to be in the mosaic.  Such theories still retain some novel theoretical virtues: they make statistical mechanical probabilities unnecessary and they are more unified and less vague than their wave-function counterparts.}  

Is the new theory more likely to be true (or close to truth, in the relevant domain) than its competitors? Given its empirical equivalence to many other theories including the quantum Mentaculus, we have to appeal to super-empirical virtues to settle the question. It is unlikely, however, we will ever arrive at a definitive answer. Nevertheless, we can develop each theory in detail, try to understand its implications, and be clear about its costs and benefits.   Only after that can we meaningfully compare the competing theories side by side as complete packages. 

We have seen that  the Humean interpretation of the Wentaculus   leads to new insights into the nature of time asymmetry and quantum entanglement. Such insights can be useful  to Humeans and non-Humeans alike.


\section*{Acknowledgement}

I am grateful for helpful discussions with  David Albert, Craig Callender, Jonathan Cohen, Chris Dorst, Michael Esfeld, Sheldon Goldstein,  Veronica Gomez, Ned Hall, Matthias Lienert, Barry Loewer, Tim Maudlin, Elizabeth Miller, Jill North, Ezra Rubenstein, Jonathan Schaffer,  Ted Sider, Roderich Tumulka, and David Wallace. I would also like to thank the participants of the  Humean Laws Workshop (July 2020) for thoughtful comments.



\bibliography{test}


\end{document}